\documentstyle[12pt]{article}
\input epsf.tex
\newcommand{\ncm}{\newcommand}
\ncm{\rencm}{\renewcommand}
\rencm{\thefootnote}{\mbox{\protect{$\fnsymbol{footnote}$}} }
\begin{document}

\begin{flushright}
TIFR/TH/95-03 \\
hep-lat/9501036 \\
\end{flushright}

\bigskip
\bigskip

\begin{center}
\large{ LANDAU GINZBURG MODEL  AND DECONFINEMENT TRANSITION 
FOR EXTENDED $SU(2)$ WILSON ACTION } \\

\bigskip
\bigskip
\large { Manu~ Mathur\footnote {On leave from and address after 
January 1995 : S. N. Bose National Centre 
for Basic 
\phantom{Scie}Sciences, DB-17, Salt Lake Calcutta-700 064, India.\\ 
\phantom{alcu}e-mail: mathur@theory.tifr.res.in} 

\bigskip

Theoretical Physics Group \\
Tata Institute of Fundamental Research \\
Homi Bhabha Road, Bombay 400 005, India} \\
\end{center}
\bigskip
\begin{center}
\bigskip

\vspace{1.0cm}

\hrule 
\vspace{0.3cm}

{\bf ABSTRACT}\\ \end{center}
\noindent  We compute the effective action in terms of the Polyakov loop 
for the 3-dimensional pure fundamental-adjoint $SU(2)$ lattice gauge 
theory at non-zero temperatures using the strong coupling expansion. 
In the extended coupling plane we show the existence of a 
tricritical point where the nature of the deconfinement transition 
undergoes a change from second to first order. The resulting phase 
structure is in excellent agreement with the Monte Carlo results both 
in the fundamental and adjoint directions. The possible consequences of 
our results on universality are discussed. \\ \\ 

\hrule
\newpage
\begin{center}
\large {\bf 1. INTRODUCTION} \\
\end{center}
\bigskip

The deconfinement transition in  gauge theories is well known
to be intimately connected to the spontaneous symmetry breaking 
of the corresponding center symmetry with the Polyakov loop ($L$) 
as the order parameter. As a consequence of universality, various 
 gauge models in $(d+1)$ dimensional Euclidean space  get related 
to much simpler ($d$) dimensional scalar or spin models near their 
critical fixed point \cite {Sve,SveYaf}.  The confining and deconfining  
phases in the language of spin models correspond to 
disordered ($<L>=0$)  and ordered phases ($<L> \neq 0$), respectively. 
The effective action written 
in terms of the Polyakov loop, after integrating out all the other degrees 
of freedom, contains all the  relevant information about the phase structure 
of the theory. These effective actions for SU(N) gauge theories and their 
universality with  spin models have been well studied in the past using 
analytical tools like large N \cite{DamPat,GocNerRos,BilCasDadMagPan}, 
mean field \cite{DroJurKrz,Ogi},  strong coupling \cite{PolSzl,GreKar} and 
weak coupling \cite{Wei} approximations as well as by Monte Carlo 
techniques \cite{DroJurKrz,DamHas}.

On the other hand, regulating the theory on a lattice,  it is expected that 
the various choices of gauge invariant lattice actions, which 
in the naive classical continuum limit (i.e $ \hbar \rightarrow
 0, a \rightarrow 0$) reduce to the Yang Mills theory  
 also  correspond to working within the same universality class. 
However, in the absence of full non-perturbative renormalization group 
equations, different lattice models expected to be in the same universality 
class have been explicitly tested for universality and their scaling 
behaviours explored both  analytically and  by Monte Carlo simulations. 
A class of lattice models, with a rich  phase structure and therefore 
well studied in the literature is  the pure fundamental-adjoint 
$SU(N)$ Wilson action  defined by the partition function 

\begin{eqnarray}
Z = \int \prod_{n,\mu} dU_{\mu}(n) \exp  \sum_P \left[{\beta_F \over 2} 
Tr_F U_P + {\beta_A \over 3} Tr_A U_P \right]
\label{AA}
\end{eqnarray}

Here $F,A$ denote fundamental and adjoint representations, respectively, 
and $U_P$ is the ordered product of the four directed link variables
$U_\mu(n) (\in SU(N))$ which form an elementary plaquette.  The sum over $P$ 
denotes the sum over all independent plaquettes of the lattice. $\int \prod_
{\mu,n}dU_{\mu}(n)$ is the product of group invariant Haar measure over all 
the links of the lattice. 
This model represents a particular choice of lattice action from a more 
general class defined by summing the above single plaquette action over 
all the representations of the gauge group with corresponding couplings.  
As in the naive classical continuum limit, the theory is only sensitive 
to the Lie algebra, thus the above class  is believed to be in the 
same universality class as that of Wilson action.  
In fact, the universality between (\ref{AA}) and the Wilson action 
({\it i.e}, $\beta_A=0$) has been explicitly demonstrated in perturbation 
theory  near the critical fixed point \cite{GonKor}, and, as expected, 
the two differ only in their scale factors $(\Lambda)$.

In the special case of SU(2), the phase structure and the scaling properties 
of the above model  have been extensively 
studied  by Monte Carlo simulations \cite{BhaCre,BhaDas,GavKarSat}  as well as 
by analytical  methods \cite{GonKor,BroKesLev,DasHelNeu,CanHalSch,AlbFlyLau}. 
Along the $\beta_F=0$ axis, it describes the SO(3) gauge theory 
with a first order phase transition at $\beta_A^c \sim 2.5$. At 
$\beta_A=\infty$, it corresponds to the $Z_2$ lattice gauge theory
with a first order phase transition at $\beta^c_F={1\over2} \ln(1+\sqrt{2}) \sim
0.44$.  From the $\beta_F=0$ and $\beta_A=\infty$ axes, the above two bulk 
transitions extend into the $(\beta_F,\beta_A)$ plane, meet at a point  
and then continue as another line of bulk first order phase 
transitions. The fact that the latter line ends at a point in the phase 
diagram allows one to bypass all the bulk singularities without losing 
confinement in $SU(2)$ gauge theory.
This model therefore retains its strong coupling 
confining property also in the continuum and is expected to be in the 
same universality class as that of the Wilson action $(\beta_A = 0)$.  

However, as the above phase diagram was mainly established using Monte Carlo 
simulations on relatively small lattices, we recently argued \cite{us1,us2,us3} 
that it is incomplete and besides the bulk transitions discussed above,
one must also worry about deconfinement phase boundary in the extended 
coupling $(\beta_F,\beta_A)$ plane. Monitoring the Polyakov loop 
expectation values $<|L|>$ and its susceptibility 
by Monte Carlo techniques, it was observed that the known second order 
deconfinement transition at $\beta_A=0$ moves into the phase diagram as  
$\beta_A$ is increased and eventually joins  the first order bulk 
transition line  at its endpoint around $\beta_A=1.0$. The above merging  
together of the deconfinement transition line with the previously claimed 
bulk transition line lead to a paradox as their nature as well as scaling 
behaviours are completely different. Even more surprisingly, it was 
observed that  

\begin{enumerate}

\item For  $\beta_A  \le 1$,  the model was  in the universality 
class of the 3-dimensional Ising model as predicted by the universality 
hypothesis \cite{Sve,SveYaf,EngFinWeb}.
However, the deconfinement transition  was found to be ${\it discontinuous}$  
above $\beta_A \approx 1$ showing a definite {\it qualitative} change in 
the properties of the theory and  the existence of a tricritical point. 

\item Absence of the  previously claimed  first order bulk transition. There 
was good evidence that  the above  transition was 
misidentified in the Monte Carlo simulations and is, in fact, first order 
deconfinement transition mentioned above. It was found 
that the first order portion of the deconfinement transition lies exactly 
along the previously known bulk transition line and there was no evidence 
of two transitions.  The plaquette susceptibility does show a peak 
along the above transition on smaller lattices, however, on going to larger 
lattices, this  peak was found to decrease considerably with the total 
4-volume of the lattice. 

\end{enumerate}

To understand the above unexpected and puzzling lattice Monte-Carlo results 
and their possible implications to the continuum physics and 
universality, one  needs to study and derive these results analytically. 
Such a study can yield further insight and a global understanding of such 
phenomena for $SU(N)$ lattice gauge theories in general. 
Analytical results also have the advantage of being free from 
finite size artifacts. 
With this motivation, in this paper 
we study the deconfinement transition and its nature in the above extended 
$SU(2)$ model in the strong coupling expansion. 
We explicitly compute the 
Landau Ginzburg effective action in terms of the Polyakov loop and show 
the existence of the tricritical point 
in the extended $(\beta_F,\beta_A)$ plane where the  the deconfinement 
transition changes its order (Fig. 2).  Moreover, the above effective 
potential reproduces the Monte Carlo phase diagram of 
\cite{us1,us2} with good accuracy (Table 1).  It is also found that within 
the strong coupling regime the tricritical point moves in the direction of 
 $(\beta_F=\infty,\beta_A=\infty)$ as $N_\tau$ increases. This possibly 
 indicates  that the continuum limit of the theory may not 
 be affected by the  existence of the tricritical point. 
Surprisingly, we find that the predictions of the critical couplings 
along the Wilson line $\beta_A=0$ 
are quite close to their corresponding Monte Carlo values (Table 2), even 
well inside the scaling region established by numerical simulations 
\cite{FinHelKar}. In the above case of pure  Wilson action,
the deconfinement transition has also been studied  by Polonyi and 
Szlachanyi \cite{PolSzl} using strong coupling expansion. However, 
our techiniques differ from theirs and our results in this special case 
$(\beta_A=0)$ are different and closer to the Monte-Carlo findings.  Following 
the work \cite{us1,us2}, a more recent paper \cite{BluTarHelKarRumTou}, 
studied the phase diagram of 
the fundamental-adjoint model for $SU(3)$ lattice gauge theory at 
nonzero temperature by Monte Carlo simulations. They reported results 
consistent with  the usual universality picture. We will briefly summarise 
and discuss their results at the end.  
\bigskip

\begin{center}
\large { 1. THE LANDAU GINZBURG MODEL } \\
\end{center}
\bigskip

We want to compute the partition function in terms of the Polyakov loop 
after integrating out all the spatial link degrees of freedom in (\ref{CC}). 
As usual, we impose periodic boundary conditions in the temperature 
direction and assume its length to be $N_\tau$. 
Defining ${\cal{U}}^{\tau}(\vec{n})=\prod_{n_0=1}^{N_\tau}U_{0}(n_0,\vec{n})$, 
the Polyakov loop can be written as, 

\begin{eqnarray} 
L(\vec{n})={1 \over 2}Tr {\cal{U}}^{\tau}\left(\vec{n}\right)
\label{CC}
\end{eqnarray}

 The  partition function (\ref{AA}) can be trivially written as 
 product over the plaquettes,  

\begin{eqnarray}
Z= \left[\prod_{Links}\int dU_{\mu}(n)\right]\prod_{Plaquettes(p)}
\left(1+G_p^F\right)\left(1+G_p^A\right),
\label{DD}
\end{eqnarray}

where, 
\begin{eqnarray}
G_p^F = \sum _{m_{p}=1}^{\infty} {1 \over m_p!} \left({\beta_F \over 2} S_p\right)^{m_p} 
\label{II}
\end{eqnarray}
and
\begin{eqnarray}
G_p^A = \sum _{n_{p}=1}^{\infty} {1 \over n_p!} \left({\beta_A \over 3} S_p^2\right)^{n_p}.  
\label{FF}
\end{eqnarray}

Here $S_p$ denotes the standard one plaquette action for the gauge group 
$SU(2)$. 

The computation of the effective action in terms of the Polyakov loops 
get considerably simplified by noticing  that after integration over 
the spatial links in an arbitrary cluster of plaquttes in (\ref{DD}), 

\begin{enumerate}
\item Due to gauge invariance of the partition function 
and the periodic boundary conditions, 
the non-trivial dependence on $U_0(n_0,\vec{n})$  
will come only if the cluster contains at least one set of plaquettes 
forming  closed loops in the  temporal direction. Moreover, the above 
dependence can come only implicitly through the Polyakov loops\footnote{
A generic gauge invariant term constructed out of $U_{0}(n_{0},\vec{n})$ 
at spatial site $(\vec{n})$ is $Tr\left({\cal{U}}^\tau\left(\vec{n}\right)
\right)^{m}$ for some integer m.  After making a gauge  transformation  
to diagonalise ${\cal{U}}^{\tau}(\vec{n})$, the above term 
can be written in terms of $L(\vec{n})$.}

\item The rest of the cluster's contribution to the partition function is 
either zero or proportional to some powers of fundamental or 
adjoint couplings. 
\end{enumerate} 

Therefore, after the above spatial link integrals are performed the Landau 
Ginzburg effective action $S_{eff}$ can be  defined as, 

\begin{eqnarray}
Z = \int \prod_{\vec{n}} d{\mu}(L(\vec{n}))\exp - S_{eff}(L). 
\label{seff}
\end{eqnarray}

Here $d\mu(L(\vec{n}))$ is the effective measure in terms of the Polyakov
loop and is computed below. 
For matter of convenience and clarity, we 
also define, 

\begin{eqnarray}
S_{eff} \equiv - \sum_N \sum_{M_F} \sum_{M_A} S_{[N,M_F,M_A]}
\left(\beta_F,\beta_A,L\right) 
\label{GG}
\end{eqnarray}

Here  N is the number of the closed loops in the temporal direction 
({\it e.g}, in Fig. 1 N=2, 4 and 6 respectively). For $SU(2)$, it is an 
even integer implying the invariance under the $Z_2$ center symmetry. 
$S_{[N,M_F,M_A]}\left(\beta_F,\beta_A,L\right)$ is a functional of 
the Polyakov loop $L(\vec{n})$. 
$M_F$ and $M_A$ are integers and  denote the highest powers of $\beta_F$ 
and $\beta_A$ in $S_{[N,M_F,M_A,L]}$. 
As we are interested 
in investigating the presence of a tricritical point in the above theory, 
we compute  the polynomial $S_{[N,M_{F},M_{A}]}(\beta_F,\beta_A)$ 
for N=2,4,6 in the strong coupling expansion.  
The higher order terms in $L$ are found to be negligibly small in the 
region of interest in the extended coupling plane and do not play any 
significant role in determining the phase boundary  and the nature of 
the deconfining transition and therefore will be ignored here onwards. 

As mentioned earlier, the measure over the Polyakov loop can be fixed by 
the gauge invariance and periodic boundary conditions at finite temerature. 
We define,

\begin{eqnarray}
{\cal{U}}^{\tau}(\vec{n}) = {\cal{U}}_0^{\tau}(\vec{n}) \sigma_0 
+ i \sum_{i=1}^{3}{\cal{U}}_i^{\tau}(\vec{n})\sigma_{i}
\label{EE} 
\end{eqnarray}

In the above equation, $\sigma_{0}$ and $\sigma_{i}$  are identity and the 
Pauli matrices respectively.
Now at a particular spatial site $(\vec{n})$, the measure over temporal 
links, $\prod_{n_0=1}^{N_\tau}\int dU_0(n_0,\vec{n})$, can  be written as 
$\left[\prod_{n_0=2}^{N_\tau}\int dU_0(n_0,\vec{n})\right]
\int d{\cal{U}}^{\tau}(\vec{n})$. As the integrand at site $(\vec{n})$ 
depends only on ${\cal{U}}^{\tau}_0 (\vec{n})$, the first $(N_\tau-1)$ Haar 
integrals can be trivially done and the integrals over 
${\cal{U}}^{\tau}_i (\vec{n})$ give the Jacobian ${\cal{J}}(\vec{n})$

\begin{eqnarray}
{\cal{J}}(\vec{n})= \left({\pi \over 2}\right)\left(1-L^2(\vec{n})\right)
^{1 \over 2} 
\label{Jac}
\end{eqnarray}

\vspace{0.2cm}

Here the factor ${\pi \over 2}$ in  the measure is due to 
the normalisation $(\int dU =1)$. 

\vspace{0.4cm}

The leading strong coupling diagrams contributing 
to the effective action are shown in Fig. 1-a,b,c. We define $\gamma_F 
\equiv {1 \over 2!} \left({\beta_F 
\over 2} \right)^2$ and $\gamma_A \equiv {\beta_A \over 3}$.  After some 
extensive computations \footnote{The details  will be presented elsewhere.} 
for  $N_\tau > 2$, the leading local contributions to the effective action 
are:  

\begin{eqnarray} 
&S_{[0,4,2]}& =\Bigg(\gamma_F + \gamma_A + \gamma_F\gamma_A -{1 \over 3!}\left(\gamma_F\right)^2 
+ {1 \over 2!} \left(\gamma_A\right)^2\Bigg) N_{p} \\
\label{V0}
&S_{[2,N_\tau,0]}& =4 \left({\beta_F \over 4}\right)^{N_\tau} \sum_{\vec{n},i}L(\vec{n})L(\vec{n}+i)\\ 
\label{L2}
&S_{[4,2N_{\tau},N_{\tau}]}& = \left(\gamma_F +\gamma_A\right)^{N_{\tau}} 
Y\left(L(\vec{n}),L(\vec{n}+i)\right) \\
\label{L4}
&S_{[6,3N_{\tau},N_{\tau}]}&= \left(\left({1 \over 3!}\right)\left({\beta_F 
\over 2}\right)^3+{1 \over 2}\gamma_A \beta_F\right)^{N_{\tau}} 
Z\left(L(\vec{n}),L(\vec{n}+i)\right)
\label{L6}
\end{eqnarray}

\vspace{0.5cm}

In the above equations we have used the notation:

\begin{eqnarray}
&Y\left(L(\vec{n}),L(\vec{n}+i)\right)& \equiv \Bigg({1 \over 3}\Bigg)^{N\tau}\sum_{\vec{n},i}
\Bigg(16 L^{2}(\vec{n})L^{2}(\vec{n}+i) -8L(\vec{n})+4\Bigg)\\
\label{Y}
&Z\left(L(\vec{n}),L(\vec{n}+i)\right)& \equiv  \left({1 \over 3}\right)^{N\tau}\sum_{\vec{n},i}\Bigg(32 
L^{3}\left(\vec{n}\right)L^{3}\left(\vec{n}+i\right)+8L\left(\vec{n}\right)L\left(\vec{n}+i\right)\nonumber\\
&&~~~~\left(1-2\left(L^{2}\left(\vec{n}\right)+L^{2}\left(\vec{n}+i\right)\right)\right)+4\Bigg)
\label{Z}
\end{eqnarray}

\vspace{0.3cm}

The local higher order corrections to the above leading effective action 
come from filling 
the diagrams in Fig. 1. by extra plaquettes.  The two and four plaquettes
corrections to $L^2$ and $L^4$ terms\footnote{In \cite{PolSzl}, $\beta_A=0$ 
and the plaquettes correction $L^2$ term in the effective action is computed 
upto order $\gamma_F$, {\it i.e}, the  first term in (16). However, its 
coeffecient is $\gamma_F$ instead of $-{1 \over 3} \gamma_F$ in our case.}
are, 

\begin{eqnarray}
&S
_{[2,N_\tau+4,2]}& =  \Bigg(-{1 \over 3}\gamma_F + \gamma_A - {{\left(N_\tau+2\right)} \over 3} \gamma_F\gamma_A
+ 2 \left({1 \over 3}\right)^2 \left(N_\tau+2\right) \left(\gamma_F\right)^2 \nonumber \\
&&~~~~~~~+{1 \over 2!} N_\tau \left(\gamma_A\right)^2\Bigg) N_{\tau} S_{[2,N_\tau,0]} \\
\label{L224} \nonumber \\
&S_{[4,2N_{\tau}+2,N_{\tau}+1]}& = N_{\tau} \left(\gamma_F+\gamma_A\right)^{\left({N_{\tau}}-1\right)}
\left(\left(\gamma_F+\gamma_A\right)^2-\left(\gamma_F\right)^2\right)\nonumber \\
&&~~~~~~Y\left(L(\vec{n}),L(\vec{n}+i)\right) 
\label{L42}
\end{eqnarray}

\begin{eqnarray}
S_{[4,2N_{\tau}+4,N_{\tau}+2]}&=& \Bigg[\left({9 \over 4}\right){}^{N_\tau}C_{2}\left(\left(\gamma_F
+\gamma_A\right)^2 -\left({2 \over 3}\right)
\left(\gamma_F\right)^2\right)^2\left(\gamma_F+\gamma_A\right)^{{N_\tau}-2}\nonumber \\
&+&\left({3 \over 2}\right){}^{N_\tau}C_{1}\left(\left(\gamma_F+\gamma_A\right)^3-\left({14 \over 15}\right)
\left(\gamma_F\right)^3 -2\gamma_F^{2}\gamma_A\right)\left(\gamma_F+\gamma_A\right)^{{N_\tau}-1}\nonumber \\
&-&{N_\tau}\left(1+{3 \over 2} N_\tau\right)\left(\left(\gamma_F+\gamma_A\right)^2 -\left({2 \over 3}\right)
{\gamma_F^2}\right)\left(\gamma_F+\gamma_A\right)^{{N_\tau}}\nonumber \\
&+&\left({1 \over 2}\right) N_{\tau}\left(N_{\tau}+1\right)\left(\gamma_F+\gamma_A\right)^{N_{\tau}+2}\Bigg]
Y\left(L(\vec{n}),L(\vec{n}+i)\right)
\label{L44}
\end{eqnarray}

\vspace{0.5cm}

For $N \ge 4$ in (\ref{GG}),  the strong coupling diagrams also give 
non-local contributions to S. However, these contributions are 
proportional to much higher powers of the fundamental and adjoint 
couplings  and are  small near the investigated phase boundaries 
and thus have been ignored.  
The Landau Ginzburg effective potential can be trivially 
obtained by putting $L(\vec{n}) = L$ = constant in the above effective 
action.  The effective potentials for $\beta_A$ = 0.5,0.75,0.9,1.4,1.5 and 
1.75 at $N_\tau=4$ are plotted in Fig. 2-a,b,c,d,e,f. In  Fig. 2-a,b,c 
corresponding to a second order deconfining transition, the 
three effective potential curves are at $\beta_F^{critical}$ and 
$\beta_F^{critical} \pm 0.05$. In the Fig. 2-d,e,f, where the transition 
becomes first order, the potentials are at $\beta_F^{critical}$ and 
$\beta_F^{critical} \pm 0.008$.  They  clearly show a 
dramatic change in the physical properties of the theory for large 
values of the adjoint couplings. Both in the second and first order 
regions the sharpness of the transition increases with increasing 
values of the adjoint coupling as found earlier by Monte Carlo simulations 
\cite{us1}. Moreover, in the first order transition region,   
the discontinuity in the Polyakov loop is large, leading to   strong 
first order transitions. In fact, at $\beta_A=1.5$ the Monte Carlo value 
of $<|L|>$ is 0.5 \cite{us1}, in close agreement with its analytical 
value 0.55 from Fig. 2-e. 
The predicted  $\beta_F^{critical}$ for the various values of the adjoint 
couplings $\beta_A$ at $N_\tau=4$ along with their corresponding Monte-Carlo 
values \cite{us1,us2} are given in Table 1. In all cases good agreement
is found between the two. The tricritical points for $N_\tau=4$ and 
$N_\tau=6$ are 
found to be at $[\beta_F=1.352, \beta_A=1.296]$ and  
$[\beta_F=1.6711,\beta_A=1.864]$, respectively. 
For $N_\tau=2$,  the above contributions to the effective potential are 
slightly different and are presently under investigation. However, both Monte
Carlo simulations and lower order effective action indicate that the 
tricritical point 
is in the vicinity of $\beta_A \approx 0.7-0.8$. The above strong coupling 
shifts in the direction $(\beta_F=\infty,\beta_A=\infty)$ 
is probably  an indication that the continuum limit is not affected by the 
existence of the tricritical point. However, such a contrained movement 
of the tricritical point with $N_\tau$, if also true in the intermediate 
and weak coupling 
regions, requires analytical explanation. The more interesting possibility 
of the existence of a new continuum theory in the extended coupling plane 
is still an open question.  It will also be interesting to investigate 
other quantitative properties $\left(e.g, ~ {T_c~ \over \sqrt{\sigma}}
\right)$ of this model for higher values of $\beta_A$ by Monte Carlo 
simulations \cite{Gav}. This requires very large 
lattices because of the known strong violation of the scaling relation in 
the large $\beta_A$ region \cite{GonKorPeiPer,GavKarSat}. 
However, the more important problem is to understand the origin for such 
a drastic change in the qualitative properties of the theory on the lattice 
itself and  to identify the degrees of freedom and the mechanism which make 
the deconfining transition first order.
An obvious guess for the above phenomenon is the different global 
properties of the $SU(2)$ and $SO(3)$  
groups. This difference can be formally  described by writing the $SO(3)$ 
part of the action (\ref{AA}) in its Villain form \cite{CanHalSch}. In the 
latter model the above difference corresponds to the $Z_{2}$ vortices 
associated with plaquette fields, $\sigma_{p} = \pm 1$ in the action. 
In the extreme case, when all the vortices are suppressed the 
above Villain action  reduces to the Wilson action  with modified coupling 
with only a second order transition.  Therefore, it is expected that the 
above change in  the qualitative properties  is due to the condensation of the 
vortices. A more careful study  both via strong coupling 
as well as Monte-Carlo simulations is possible by controlling these objects 
with a potential   term  $\lambda \sum_{plaquettes} \sigma_p$ in the action 
and studying its effect on the deconfinement transition. It will be  
interesting to see the dependence of the tricritical point on $\lambda$.
Work in this direction is in progress and will be reported elsewhere.

Table 2 gives the predicted critical couplings and their Monte Carlo 
values \cite{FinHelKar} along the Wilson line $(\beta_A=0)$ at different 
$N_\tau$. It clearly shows that the predictions of the strong coupling 
analysis for the deconfining transition are very close to the corresponding 
Monte Carlo values up to $N_{\tau}=16$. Further comparisons in the region even 
closer to the asymptotic scaling region could not be made because of the lack 
of Monte Carlo data. The above validity of the strong coupling expansion  
in the scaling region $(N_\tau \ge 4)$ found by Monte-Carlo simulations
\cite{FinHelKar} and  beyond $\beta_F=2.2$, where the crossover 
from strong coupling to weak coupling is expected, is again a surprising 
phenomenon.  However, it  needs further verifications with other lattice 
observables and a careful study of the crossover region at $T \neq 0$.

In  the forementioned recent  paper \cite{BluTarHelKarRumTou} studying 
$SU(3)$ deconfinement transition at non-zero temperature, no evidence 
for the change in the order of the deconfining transition was reported. 
 Also by extrapolating the Polyakov loop data  to infinite volume, the 
 deconfinement transition was seperated from the  discontinuity in the 
 plaquette corresponding to the bulk transition. 
However,  scaling laws  for the plaquette discontinuity remained untested. 
The  discrepancies in the qualitative aspects of the phase diagrams of 
SU(2) and SU(3) lattice  gauge theories need an explanation. One should again 
check the role of $Z(3)$ vortices in $SU(3)$ deconfinement transition. 
In \cite{us2} the $SU(2)$ plaquette susceptibility peak was found to 
decrease on going to relatively large volume. Therefore,  it is 
important to to verify the scaling behaviour of the bulk transition 
in the case of $SU(3)$ also.

Finally we summarize the findings of this paper. We have analytically 
confirmed the earlier Moonte Carlo evidence for the existence of the 
tricritical point in the extended SU(2) model.  All the qualitative 
as well as quantitative features of the deconfinement transition in 
the extended coupling plane found by Monte Carlo simulations are 
reproduced to a good accuracy. As discussed above,  
this study also opens some new analytical as well as numerical avenues 
for understanding  and analysing
the new features associated with the extended model. Surprisingly, it is 
found that the strong coupling predictions for the critical couplings 
are valid well within the scaling region. 

\vspace{2cm} 

It is a pleasure to acknowledge Rajiv V. Gavai for  many useful discussions 
and earlier collaborations. I also thank   Prof. Michael Grady, SUNY, 
Fredonia, USA  for the fruitful first collaboration on this subject.   
It is also a pleasure to acknowledge Balram Rai for the various discussions.

\newpage

\begin{table}
\begin{center}
{Table 1}
\end{center}
\caption
{The values of the critical couplings $(N_\tau=4)$ at various $\beta_A$.} 
\medskip
\begin{tabular}{|c|c|c|c|}
\hline
                         &                         &                       \\
~~~~$\beta_A~~~~~~~~$ & $~\beta_F^{critical}~~ (Strong~ Coupling)~$ & $ ~~~\beta_F^{critical}~~ (Monte~ Carlo^{[1]})~~$ \\
                         &                         &                              \\
\hline \hline
0.5 & 1.793  &         1.83  \\
\hline
0.75  &  1.619   &      1.610   \\  
\hline
0.9 &  1.531     &       1.489              \\
\hline
1.1 & 1.432     &       1.327             \\
\hline
1.4 & 1.314     &        --              \\
\hline
1.5 & 1.28     &         1.05           \\
\hline
1.75 & 1.147   &          --              \\
\hline
\end{tabular} 
\end{table}
\begin{table}
\begin{center}
{Table 2}
\end{center}
\caption
{The scaling of the critical coupling with $N_\tau$ at $\beta_A=0$.} 
\medskip
\begin{tabular}{|c|c|c|c|}
\hline
                         &                         &                       \\
~~~~$N\tau~~~~~~~~$ & $~\beta_F^{critical}~~ (Strong~ Coupling)$ & $ ~\beta_F^{critical}~~ (Monte~ Carlo^{[2]})~~~$ \\
                         &                         &                              \\
\hline \hline
4 & 2.00     &            2.2986(6)             \\
\hline
5 & 2.25766  &        2.3726(45)  \\
\hline
6 & 2.38535  &        2.4265(30)  \\
\hline
8  &  2.51942   &      2.5115(40)   \\  
\hline
16 &  2.81613     &      2.7395(100)              \\
\hline
\end{tabular} 
\label{tabmas} 
\vspace{2cm}
\hrule
\vspace{0.5cm}
[1] The data taken from \cite{us1,us2}.

[2] The data taken from \cite{FinHelKar}. 
\end{table}
\end{document}